\documentstyle[emulateapj,psfig]{article}

\makeatletter

\newenvironment{inlinefigure}{%
\def\@captype{figure}%
\noindent\begin{minipage}{0.999\linewidth}\begin{center}}
{\end{center}\end{minipage}\smallskip}
\makeatother

\newcommand{\mum}{$\,\mu$m}

\def\m{$\mu$m}

\lefthead{Chapman et al.}

\begin{document}
\title{Optically faint counterparts to the {\it ISO}-FIRBACK 170\m\ population:\\
the discovery of cold, luminous galaxies at high redshift}
\author{S.\,C.\ Chapman,$\!$\altaffilmark{1}
I. Smail,$\!$\altaffilmark{2}
R.\,J.\ Ivison,$\!$\altaffilmark{3}
G.\ Helou,$\!$\altaffilmark{1}
D.\,A.\ Dale,$\!$\altaffilmark{4}
G.\ Lagache$\!$\altaffilmark{5}
}

\altaffiltext{1}{Department of Physics, California Institute of Technology,
Pasadena, CA 91125}
\altaffiltext{2}{Department of Physics, University of Durham, South Rd, Durham DH1 3LE, UK}
\altaffiltext{3}{Astronomy Technology Centre, Royal Observatory,
Blackford Hill, Edinburgh EH9 3HJ, UK}
\altaffiltext{4}{Department of Physics \& Astronomy University of Wyoming Laramie, WY 82071} 
\altaffiltext{5}{Institut d'Astrophysique Spatiale, Universite Paris Sud, Bat.\ 121, 91405 Orsay Cedex, France}

\slugcomment{Accepted by the Astrophysical Journal}

\begin{abstract}
We present Keck spectroscopy and UKIRT near-IR imaging observations of
two 170\,\m-selected sources from the {\it ISO}-FIRBACK survey which
have faint counterparts in the optical, and $r-K\sim5$.  
Both sources were expected to
lie at $z>1$ based on their far-infrared, submillimeter and radio
fluxes, assuming a similar spectral energy distribution to the local
ultra-luminous infrared galaxy (ULIRG) Arp\,220.  However, our spectroscopy
indicates that the redshifts of these galaxies are $z<1$:  $z=0.91$ for
FN1-64 and $z=0.45$ for FN1-40.  While the bolometric luminosities of
both galaxies are similar to Arp\,220, it appears that the dust
emission in these systems has a characteristic temperature of
$\sim$30\,K, much cooler than the $\sim$50\,K seen in Arp\,220.
Neither optical spectrum shows evidence of AGN activity.  If these
galaxies are characteristic of the optically faint FIRBACK population,
then evolutionary models of the far-infrared background must include
a substantial population of cold, luminous galaxies.
These galaxies provide an important intermediate comparison between the local
luminous IR galaxies, and the high redshift submillimeter-selected
galaxies, for which there is very little information available.
\end{abstract}


\section{Introduction}
\label{secintro}

%
%
\begin{figure*}[htb]
\centerline{
\psfig{figure=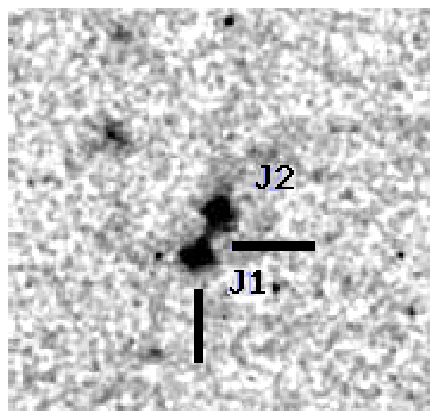,angle=0,height=1.2in} 
\psfig{figure=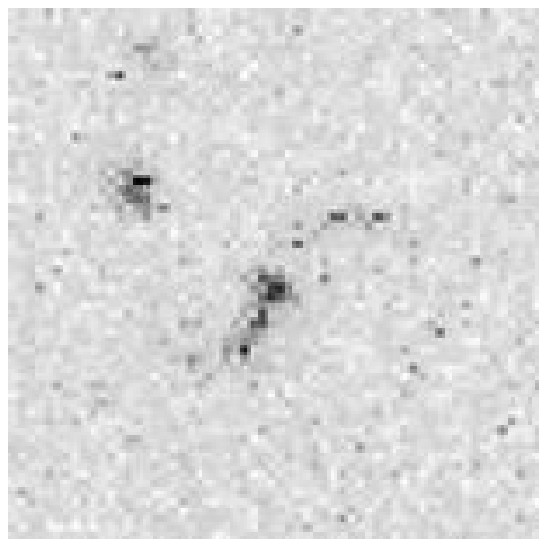,angle=0,height=1.2in} 
\psfig{figure=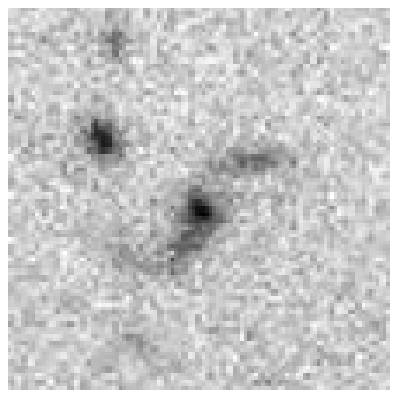,angle=0,height=1.2in} 
\psfig{figure=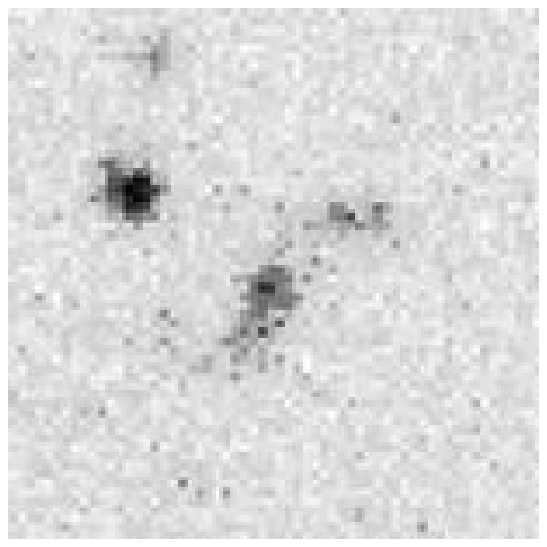,angle=0,height=1.2in} 
\psfig{figure=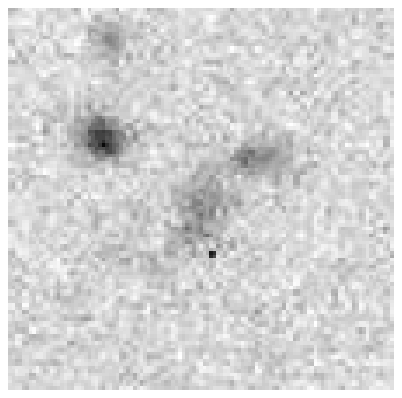,angle=0,height=1.2in} 
\psfig{figure=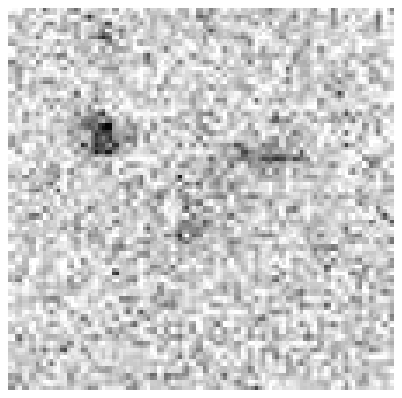,angle=0,height=1.2in}
}
\centerline{
\psfig{figure=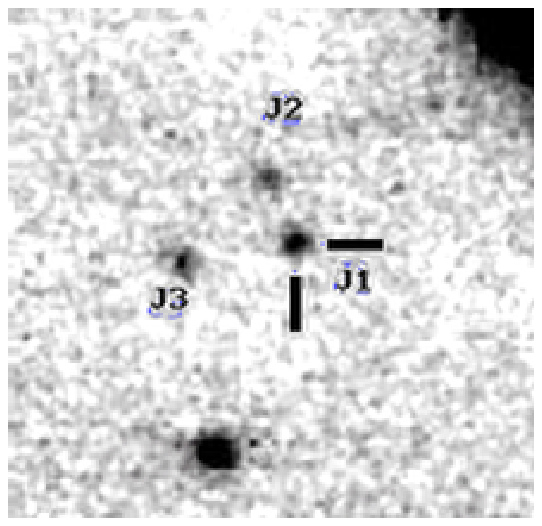,angle=0,height=1.2in}
\psfig{figure=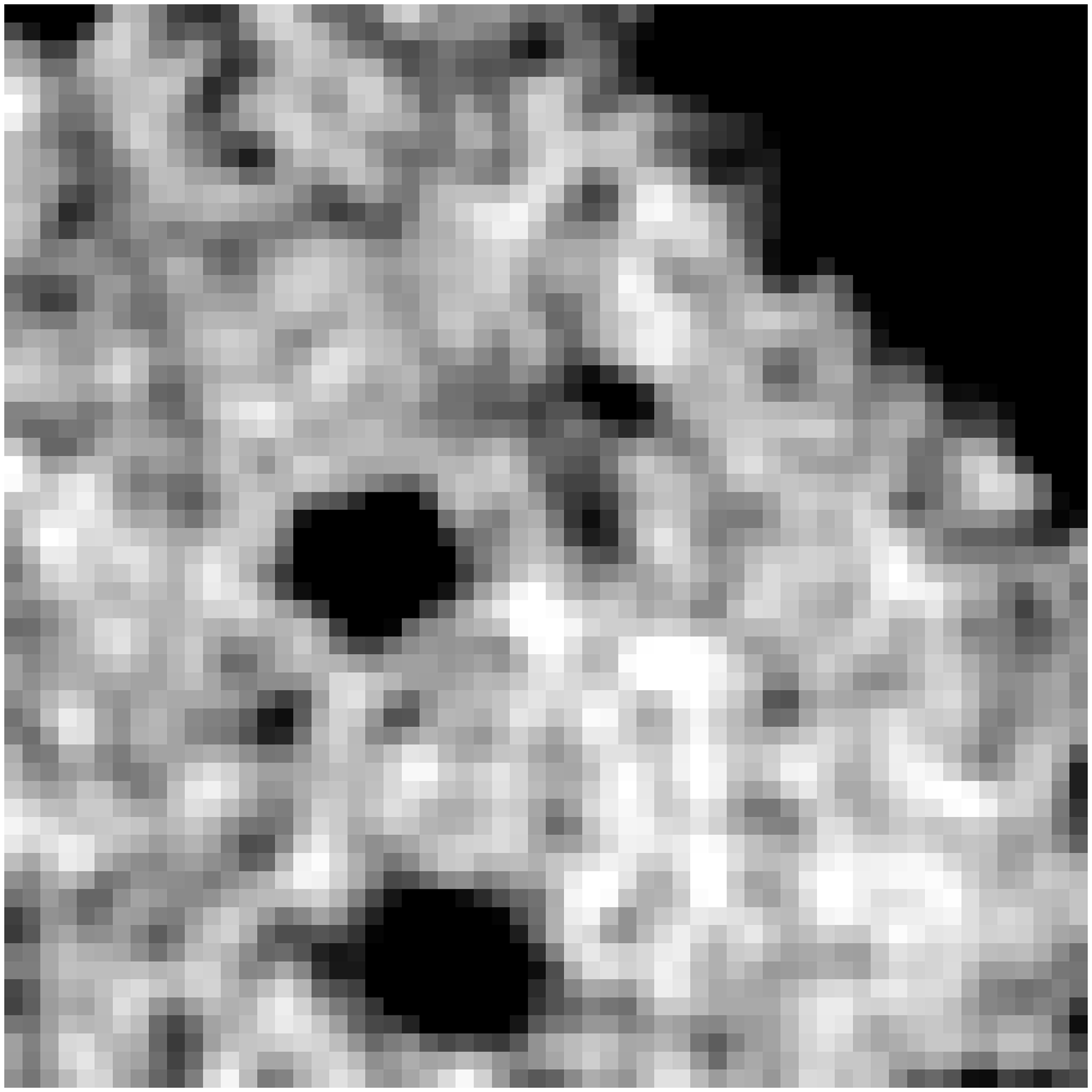,angle=0,height=1.2in} 
\psfig{figure=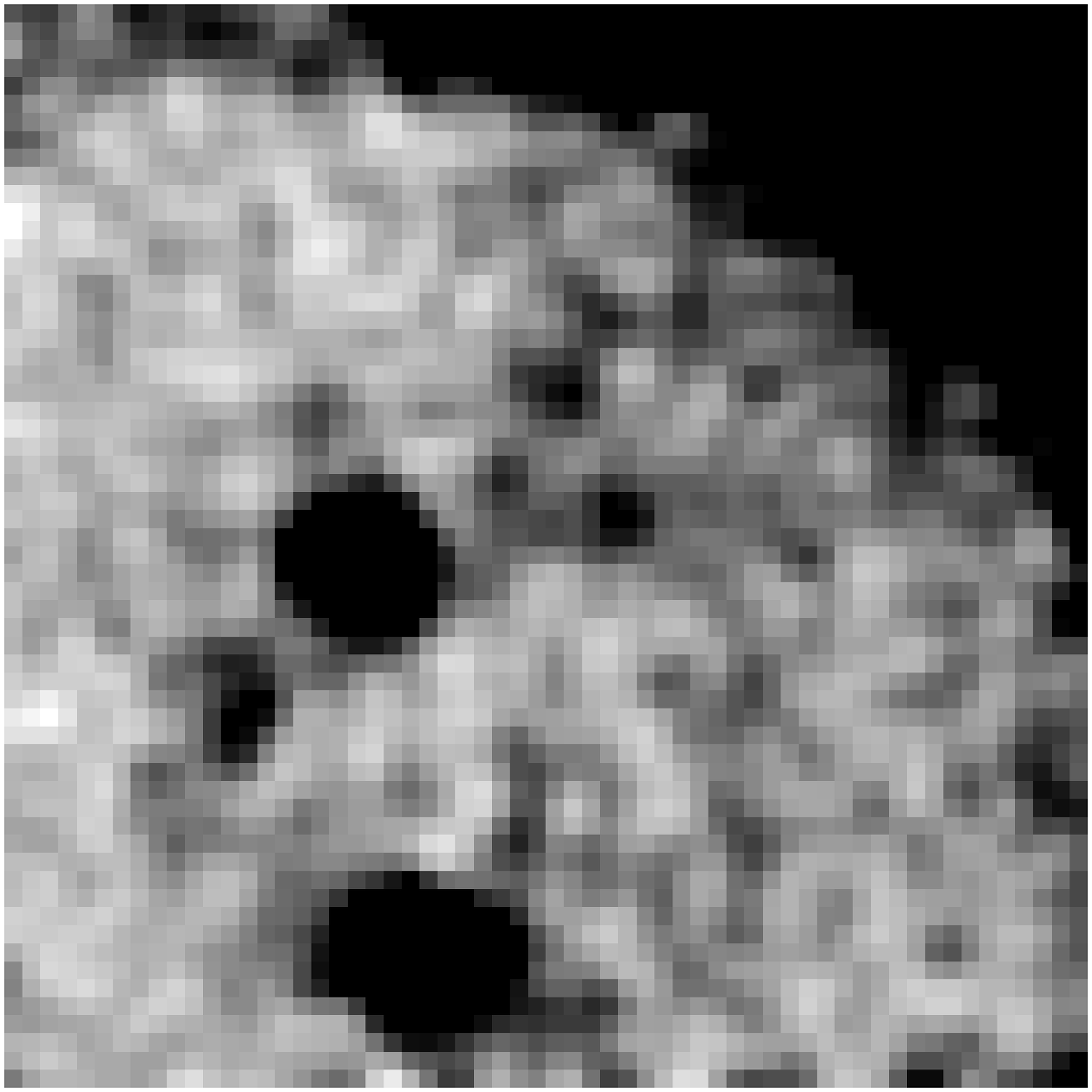,angle=0,height=1.2in} 
\psfig{figure=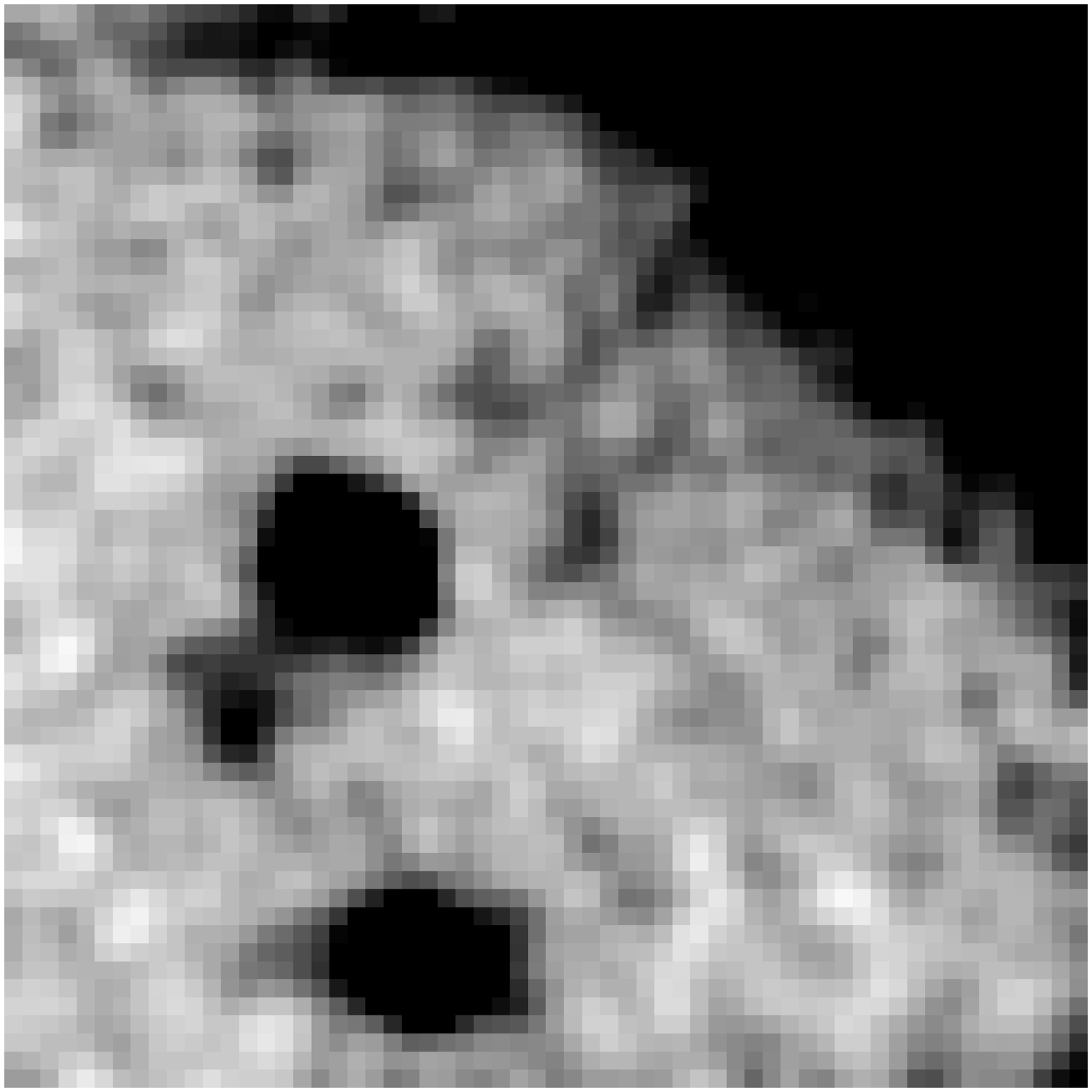,angle=0,height=1.2in} 
\psfig{figure=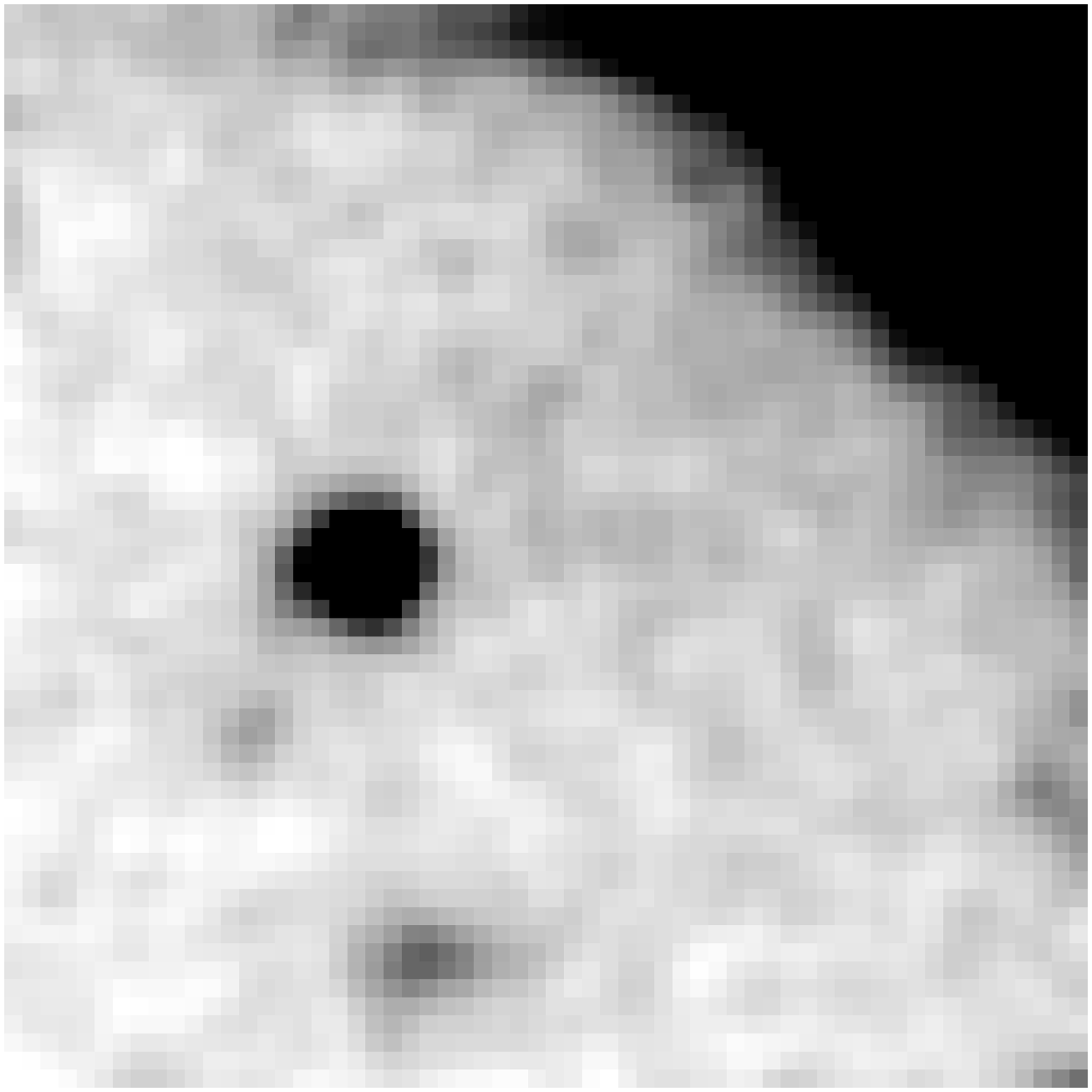,angle=0,height=1.2in} 
\psfig{figure=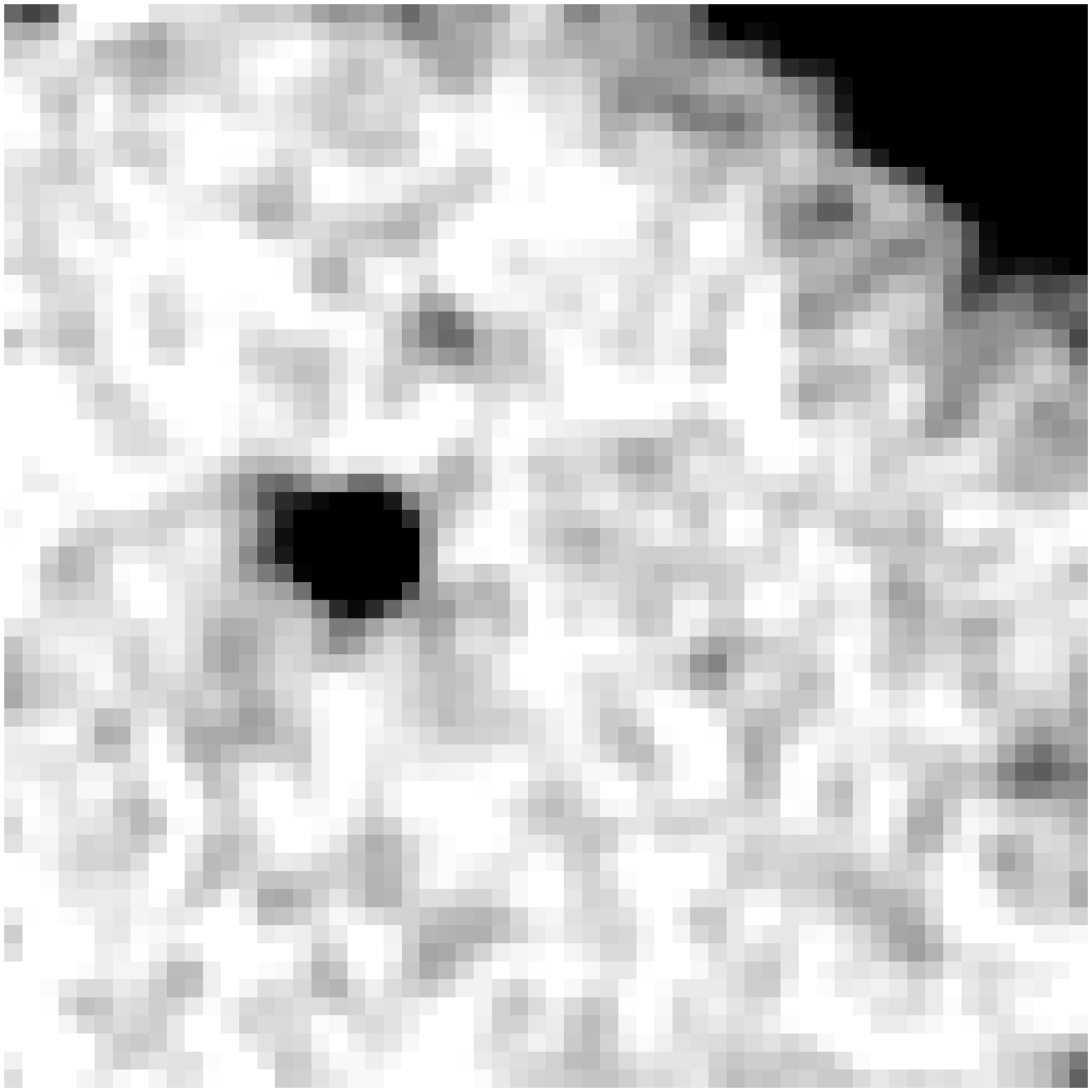,angle=0,height=1.2in}
}
\figurenum{1}
\caption{\footnotesize 
Multi-band images (from left to right are $K, z', i, r, g, U$) 
of FN1-64 ($z=0.91$ -- upper panels) and FN1-40
($z=0.45$ -- lower panels) 
used to identify the FIRBACK population (NET archive),
and our deep UKIRT
$K$-band imaging which reliably identifies the counterparts to these
two sources. The data for the fainter FN1-40 has been smoothed with the 
PSF for visibility.
The seeing at UKIRT was $0.4''$ providing a high
resolution view of these luminous dusty galaxies.  The radio source
position is shown by the cross-hair and in each case is coincident with
a very red ($r-K\sim5$) galaxy component. We identify several components in each source
(major components labelled J1, J2, etc) and conclude that both galaxies
likely show a
multi-component merging structure reminiscent of local ULIRGs.  Each
panel is $20''$ on a side with North up and East to the left.
}
\label{fig1}
\end{figure*}

The spectral shape of the far-infrared background (FIRB) detected by
{\it FIRAS} at 100\,\m--4\,mm and {\it DIRBE} at 140 and 240\,\m\ 
(Puget et al.\ 1996; Fixsen et
al.\ 1998) indicates a peak at $\sim200$\,\m\ with energy comparable to
the optical/UV background.  This peak arises from optical/UV radiation
from star formation and AGN activity in obscured galaxies at $z\gg 0$
which is absorbed by dust and reradiated in the far-infrared.  This
obscured population of galaxies could host approximately half of the
massive star formation activity over the history of the Universe (e.g.~Blain
et al.\ 1999).

Far-infrared surveys at wavelengths close to the peak of the FIRB
provide a powerful route for understanding the properties of the
obscured activity in the distant Universe and its relevance to the
formation and evolution of both galaxies and super-massive black
holes.  The FIRBACK (Far-InfraRed BACKground, Dole\ 2000)
survey obtained wide-field imaging at 170\,\m\ with the PHOT instrument
on-board {\it ISO} satellite in three separate regions of the sky
chosen for low Galactic cirrus foreground.  FIRBACK is the most
reliable and deepest ($\sigma$(170\m)$\sim 40$\,mJy) infrared census at
wavelengths near the peak in the FIRB. The FIRBACK sources down to
120\,mJy account for about 10\% of the FIRB seen by COBE at
140--240\,\m\ (Puget et al.\ 1999).  Evolutionary models
(Dole et al.~2001,
Lagache et al.~2002) suggest that
the sources identified by FIRBACK comprise both star-forming galaxies
at low redshifts, $z\sim 0.1$, and a population of much more luminous
galaxies at higher redshifts. 
The models predict that a
quarter of the FIRBACK sources should have $z>0.5$, with a tail
reaching beyond $z=1.5$, however this is quite sensitive to the depth and form
of the evolution assumed.  The sources in this high-redshift tail
provides the strongest constraints on the evolution of the population
contributing to the peak of the FIRB at $\sim 200$\m. For this reason
the identification and study of these galaxies is of particular interest
(Sajina et al.~2002, Dennefeld et al.~2002).

The coarse beam of {\it ISO} at 170\,\m\ beam produces large
uncertainties in the source positions, 
(100\arcsec\ diameter, 99\% error circle).  
However, using
the empirical radio-infrared correlation for star forming galaxies
(Helou et al.\ 1985), we can exploit deep 1.4-GHz radio data from the
VLA in C-array configuration (Ciliegi et al.\ 1999) to identify the
FIRBACK sources with positional uncertainties of only $\sim1''$ ($15''$
VLA beamsize). Most FIRBACK sources lying within the sensitive region of the
VLA image are detected in the radio (although some are not and are therefore
difficult to identify at other wavelengths). 
Approximately 80\% of the FIRBACK galaxies with radio
IDs are detected in shallow sky surveys in the optical (DPOSS2) or
near-IR (2MASS) (Dennefeld et al.~in preparation).
These 170-\m\ sources correspond to relatively low
redshift galaxies.  The remaining 20\% of the radio/FIRBACK-N1 
field population 
are identified in deep UKIRT $K$-band imaging at $17< K <21$ (Sajina et al.\
2002).
The faint near-IR magnitudes of these
galaxies support the existence of a high-redshift tail in the sample
and provides an efficient route to identifying high-redshift candidates
from the 170-\m\ population.

Additional constraints on the redshifts of the distant FIRBACK
population are provided by submillimeter observations of a subsample
of the catalog (Scott et al.\ 2000; Sajina et al.\ in preparation).
SCUBA photometry observations at 850\,\m, in conjunction with the
170\,\m\ and 1.4-GHz fluxes, provide photometric
redshift estimates (e.g.\ Carilli \& Yun 2000).  In this paper we
present Keck spectroscopy and UKIRT near-IR imaging observations of two
FIRBACK sources, FIRBACK~FN1-64 and FIRBACK~FN1-40, 
which were detected by SCUBA in the
submm (Scott et al.\ 2000) and whose far-IR/submm/radio properties
indicate they potentially lie at $z>1$, in the high-redshift tail of
the FIRBACK distribution.  Our calculations will assume a flat,
$\Lambda=0.7$ Universe, and H$_0$=65\,km\,s$^{-1}$\,Mpc$^{-1}$,
providing a scale of 8.4\,kpc per arcsec at $z=0.91$ (FN1-64) and
6.2\,kpc per arcsec at $z=0.45$ (FN1-40).

\section{Observations}

The initial identification of the FIRBACK sources involves searching
for optical counterparts in the second generation Digital Palomar
Observatory Sky Survey (DPOSS2) $F$-band images (close to gunn-$r$) of
this region, using the accurate positions of the sources from the radio
maps.  This search highlights both FN1-40 and FN1-64 as optically faint
(or undetected) sources, at the $F<22$ limit of DPOSS2.

The sources are expected to be at
$z>1$ based on photometric redshifts from the FIR/submm/radio, 
and within a degeneracy between T$_{\rm d}$ and $z$ (Blain 1999). Using the 
standard Carilli \& Yun (2000) estimator for radio/850\mum\ gives 
$z_{\rm FN1-40}=1.1$, $z_{\rm FN1-64}=1.3$, with a 1$\sigma$ uncertainty of
0.17.
Aperture matching the 
observations at different wavelengths (radio--15\arcsec, 850\mum--19\arcsec,
450\mum--11\arcsec, 170\mum--100\arcsec) could introduce errors in 
such redshift estimates, as well as SED fitting, 
with confusion problems particularly affecting the 170\mum\ band.
However, the very low source densities in the radio, 850\mum, 450\mum\ 
at the FIRBACK source flux levels preclude
any significant contamination to the beam by interlopers with a very high
probability. By contrast the beams are large enough to include
all the flux from high redshift objects.
To provide  complete, high resolution $K$-band identifications for the
FIRBACK sources a campaign is underway involving deep near-IR imaging
(Chapman et al.\ in preparation).

\subsection{UKIRT imaging}

We observed FN1-40 and FN1-64 in the $K$-band at UKIRT using the Fast
Track Imager (UFTI) as part of the on-going identification campaign of
FIRBACK sources. The source positions were taken from the radio/SCUBA
identifications as uncovered in Scott et al.\ (2000), with coordinates
listed in the catalog of Dole et al.\ (2001).  Each source was imaged
for a total of 1800\,s, with individual exposures of 60\,s each,
reaching a limiting magnitude in a $2''$ diameter aperture of $K=20.4$
(5$\sigma$) in $0.4''$ seeing.  Data were reduced using the UKIRT
software pipeline {\sc oracdr} (Bridger et al.\ 2000).

The $K$-band imaging of both sources is shown in Fig.~1.  
The precise mapping of the radio source
catalog onto the optical grid suggests the astrometric error is 
dominated by the centroiding of the radio sources themselves 
(${FWHM_{\rm 1.4 GHz}}\over{2\times S/N}$ $\sim 1$\arcsec). 
The UKIRT
observations clearly identify near-IR counterparts at the position of
the radio source in both cases. Indeed, we find several near-IR
components close to the radio counterpart, on scales of $<5''$
($<$50\,kpc at $z<1$). These components are all resolved in 0.4$''$
seeing, confirming they are all galaxies.

Based on the available optical limits we find that in both sources the
component which is coincident with the radio emission, and by
implication is the source of the luminous far-IR emission, is very red,
$(R-K)\sim5$ (Table~1).

\subsection{Archival INT imaging}

Archival, multi-band data is made publically available through the Isaac 
Newton Groups'
Wide Field Camera Survey Programme. We extracted $U, g, r, i, z'$ band
imaging for our FIRBACK galaxies, 
stacking multiple exposures in the same filters with IRAF\,{\it IMCOMB}.
The imaging has 0.33\arcsec/pixel, with an average 1.2\arcsec\ FWHM PSF.
The calibration zero points were taken from the archive listing for the
appropriate observation date ({\it www.ast.cam.ac.uk/$\sim$wfcsur/photom.html}).
Fig.~1 shows these images scaled to the UKIRT frame. 
FN1-64 displays a striking merger morphology, with components spanning a range in
SEDs. The strong $K$-band J1 component appears only as a faint tail in the optical.
FN1-40 is  a much fainter optical source, and we have smoothed the image with the PSF
for display purposes. Some faint emission between J1 ane J2 in the $I$,$z'$ images
suggests that this source may also be an interaction between multiple components.  %
We find that in both sources the
component which is coincident with the radio emission, and by
implication is the source of the luminous far-IR emission, is very red,
$(r-K)\sim5$ (Table~1).

%
\begin{figure*} 
\centerline{\psfig{figure=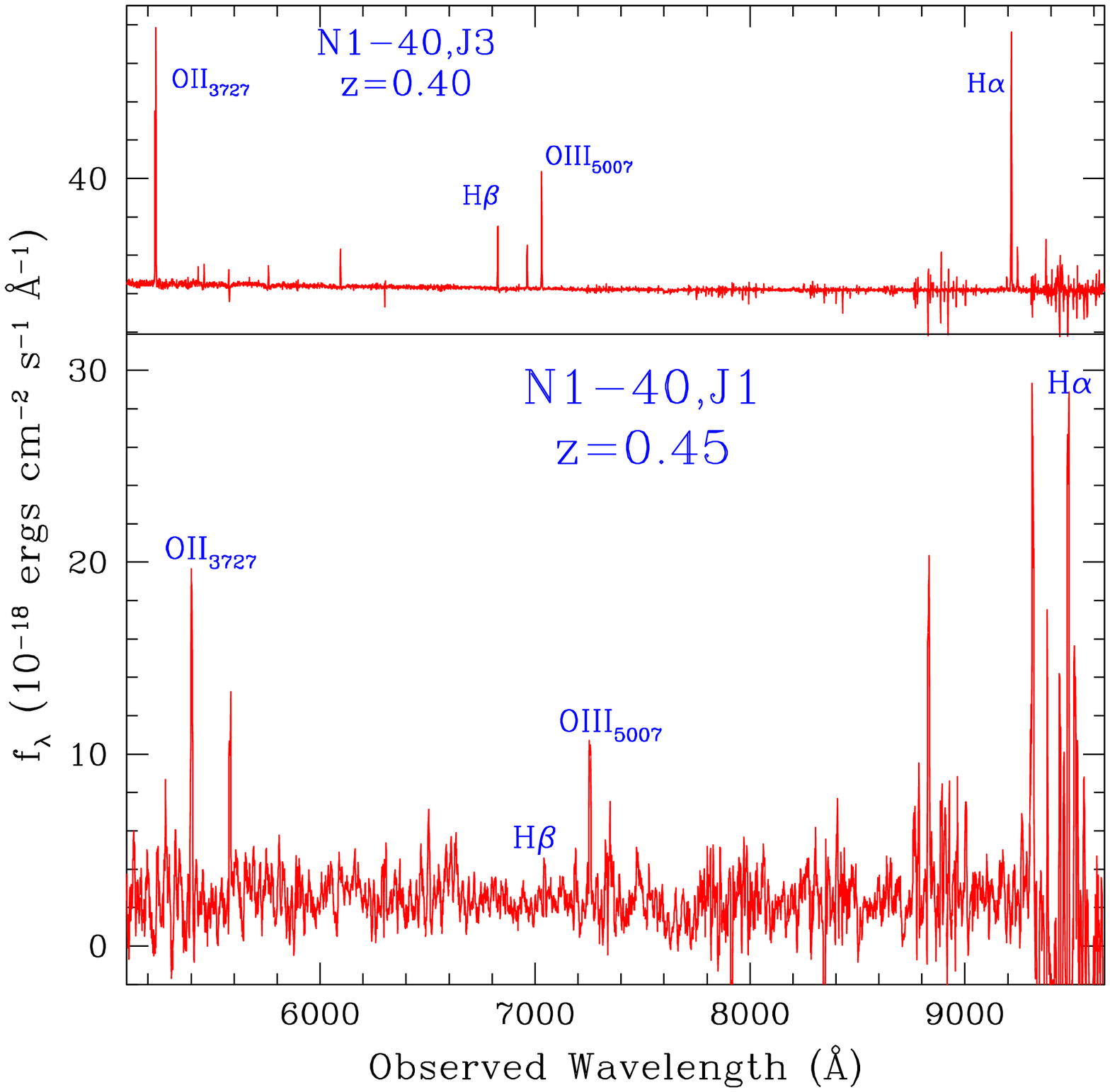,angle=0,width=3.2in}
\psfig{figure=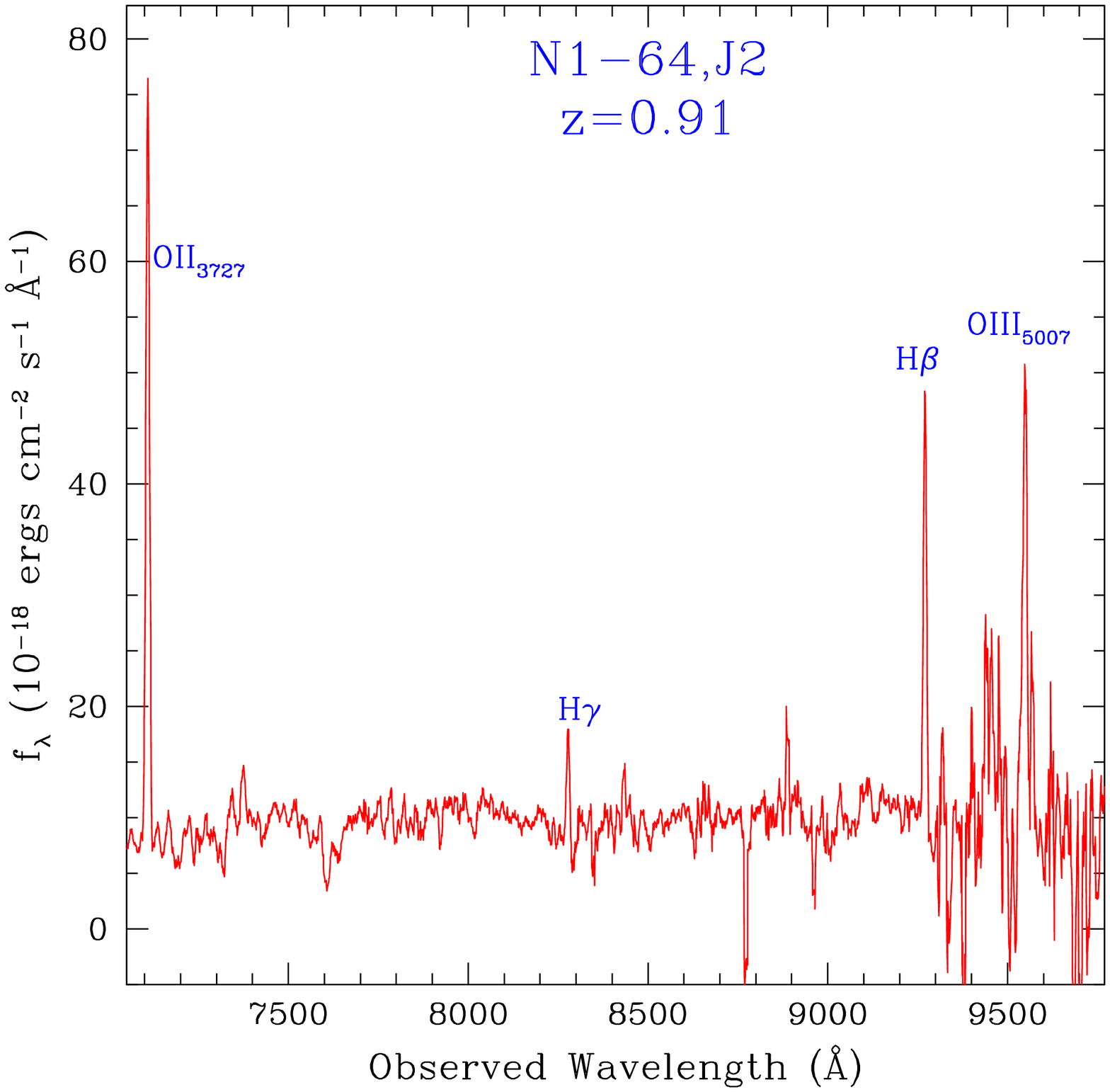,angle=0,width=3.2in}}
\figurenum{2}
\caption{\footnotesize
Keck/ESI spectra of FN1-40 (above) with the radio identified
source J1 at $z=0.45$. Inset for comparison is the optically bright
component J3 ($z=0.40$) lying on the same slit, with flux scaled down by
100$\times$.  While relatively close in redshift, the velocity difference
($\sim$13,000\,km/s) is too large for them to be part of the same
merging system.  The Balmer ratios do not show any significant
reddening.  FN1-64/J2, the optically brighter component of an obviously
merging pair, is presented below ($z=0.91$). Detected lines are
indicated. Spectral resolution is $\sim$1\AA, resolving the [O{\sc ii}]
doublet in all cases.
}
\label{fig2}
\end{figure*}

\subsection{Keck spectroscopy}

Having identified reliable counterparts to the 170-$\mu$m sources in
these two fields we are then in a position to test the prediction
that these galaxies should lie at $z>1$.

Spectroscopic observations of FIRBACK FN1-40 and FN1-64 were taken using
the Echellette Spectrograph and Imager (ESI) on the Keck\,II telescope
in 2001 July.  In the echellette mode used here, ESI provides complete
coverage of the optical waveband from the atmospheric cut-off to the
limit of the sensitivity of silicon CCDs: 0.32--1\,$\mu$m.  The high 
spectral resolution of the system, $\sim$1\AA, allows for very good
sky subtraction into the atmospheric OH forest at the red end of
the spectrum.

The galaxies were acquired and centered on the slit manually using the
optical guide camera.  In the case of FN1-40, we were able to center on
the bright galaxy, J3 (Fig.~1), with the slit aligned to cover J1, the
component identified with the radio emission.  For FN1-64, only the
optically brighter component of the merging pair, J2, was aligned on
the slit.

The spectral integrations were 1800\,s for each source.  High signal-to-noise 
flats and wavelength calibrations were taken shortly before the
observations of each target, and fluxes were calibrated using spectra
of red standard stars.  The data were reduced with the {\sc makee}
software using the standard reduction recipe outlined in the manual
(\markcite{surf}Blightman 1998).  The calibrated spectra for FN1-40/J1
and FN1-64/J2 (as well as the acquisition object J3) are shown in Fig.~2.

The spectra of both FN1-40/J1 and FN1-64/J2 show a series strong emission
lines in the red, we identify the strongest of these as [O{\sc ii}]3727,
[O{\sc iii}]5007 and for FN1-40: H$\alpha$.  Based on these identifications we
determine a redshift of $z=0.45$ for FN1-40/J1 and $z=0.91$ for
FN1-64/J2.\footnote{For completeness we note that an attempt to detect
redshifted H$\alpha$ in FN1-64/J2, using a 1800\,s $J$-band observation
with the CGS4 spectrograph on UKIRT failed to detect a line at the
relevant wavelength.} 
In addition, we measure a redshift of $z=0.40$
for the acquisition galaxy, J3, in the field of FN1-40.  More details
of the spectral line strengths from these observations are given
in Table~1.

\section{RESULTS}

The ESI spectroscopic observations provide reliable redshifts for the
proposed radio-detected counterparts to the 170-\m\ sources in the
fields of FN1-40 and FN1-64.  These redshifts, $z=0.45$ and $z=0.91$, are
below those predicted on the basis of the far-IR/submm/radio properties
of these galaxies assuming a  spectral energy distribution (SED) for these
luminous, dusty galaxies similar to that of Arp\,220.  We now
investigate this discrepency in more detail.

\subsection{Far-IR properties}

Figure~3 depicts the measured near-IR through radio photometry for the
two FIRBACK galaxies. A template fit from SED library of Dale et
al.\ (2001,2002) is overlayed, encompassing the mid-IR through radio
wavelengths.  This template library represents a parametrized sequence
of SEDs (ordered in terms of their 60/100\,\m\ restframe flux ratios)
where a range of dust properties (temperatures and emissivities) are
modelled to provide a composite SED, with increasing far-IR colors
for increasing far-IR luminosity. These SEDs provide a good 
representation of all local far-IR luminous galaxies (Dale et al.~2001,2002). 
The radio
luminosity of the model SED is tied directly to the far-IR luminosity,
using the empirical relation from Helou et al.\ (1985).

To fit these models to our observations we simply minimize $\chi^2$ 
between the SED and the observations for the radio through
far-IR points, ignoring the poor signal to noise 450\,\m\ point, and
weighting each of the points equally.  In the case of FN1-64, the
Dale et al.\ templates 
cannot simultaneously fit the radio, 850\m\ and 170\m\ bands
to within the 1$\sigma$ uncertainties, however the fit lies within the 0.2\,dex scatter of the FIR/radio relation (Helou et al.~1985).
For reference, we also show the best fit
excluding the 170\,\m\ point for illustration. The excellent resulting
fit with a slightly cooler template (lower 60/100 parameter) to the
450\,\m, 850\,\m\ and radio points suggests that the large {\it
ISO}/PHOT beam (100\arcsec\ diameter, 99\% error circle) 
may include an additional far-IR source in
addition to the one isolated by the radio and submm.
A second radio source with S$_{1.4 \rm GHz}<5\sigma$ 
(Ciliegi et al.\ 1999) 
lies towards the edge of the FIRBACK error 
circle, and may be contributing to the FIR flux.

From our fits we estimate that galaxies are relatively cool with rest frame
60/100\,\m\ ratios of 0.43$\pm$0.03 and 0.69$\pm$0.04 
for FN1-40 and FN1-64 respectively, where
flux errors lead to an uncertainty in the fit 60/100\,\m\ ratio. 
As the Dale et al.\ SED models assume a superposition of black bodies
of different temperatures, they can be described best in terms of
percentage L$_{\rm IR}$ contributions from dust at different temperatures:
T$_{\rm d}<15$\,K, $<20$\,K, $<31$\,K.  FN1-40 has respectively 42\%, 89\%,
100\%, while FN1-64 has 31\%, 78\%, 98\%.  However, for direct comparison
with other galaxies from the literature we also fit a single dust
temperature and emissivity ($\beta$) to the submm/far-IR data points
and obtain: T$_{\rm d}$=25.7$\pm$0.4\,K and $\beta$=1.76$\pm$0.01 for FN1-40; 
and T$_{\rm d}$=30.8$\pm$0.7\,K with $\beta$=1.68$\pm$0.02 for FN1-64. 
These values of $\beta$ lie within the 1$\sigma$
range of local values for IRAS galaxies, 
as fit by Dunne et al.~(2000) (median $\beta$=1.33).
We consider the effects of dust temperature further in subsequent sections.


In terms of bolometric far-IR luminosity (40 to 200\m), using the best
fitting SEDs we estimate that FN1-40 has a luminosity of around
L$_{\rm FIR}\sim 7\times10^{11}$\,L$_\odot$, placing it just in
the ULIRG category.  FN1-64 is  more luminous, with L$_{\rm FIR}\sim
4\times10^{12}$\,L$_\odot$, placing it well into the ULIRG class and showing
that it is four times as luminous as Arp\,220.  In both cases,
adopting total infrared luminosities, L$_{\rm TIR}$, covering all the
emission from 3--1100\m\ in the restframe, we estimate luminosities
which are $\sim2\times$ higher than the corresponding L$_{\rm FIR}$.

We compare in Figure~4 the rest frame 
60/100 ratios (corresponding roughly to dust temperature) 
and FIR luminosities for these FIRBACK galaxies with
a sub-sample of the {\it IRAS} Bright Galaxy Sample (BGS) with 850\mum\
measurements (Dunne et al.\ 2000). Our extrapolation to 60/100 is somewhat
model dependent; while our observed 170\m\ point measures the rest frame
100\m, our model fitting gives a variation in the 60\m\ point, dominated
by photometric uncertainties (shown with error bars in Fig.~4).
Fitting single dust temperature 
grey-bodies to local {\it IRAS}-selected galaxies using the 850\mum\ point 
(Dunne et al.\ 2000) has
demonstrated the 60/100 ratio alone to slightly overestimate T$_{\rm d}$,
with T$_{\rm d}$ showing a tighter correlation than 60/100 with L$_{\rm FIR}$.
This comparison shows that these two FIRBACK galaxies are colder than
the majority of the 
{\it IRAS} BGS given their L$_{\rm FIR}$.
However, the BGS (S$_{60\mu m}>$5.24\,Jy) 
is relatively small with very few objects as luminous as FN1-40, FN1-64.  
We would like to make a comparison with the
fainter S$_{60\mu m}>$1.2\,Jy {\it IRAS} galaxies (Fisher et
al.\ 1995).  The local 60/100\,\m\ distribution of the 1.2\,Jy sample
is characterized in detail in Chapman et al.\ (2002a).  Placing the
two FIRBACK galaxies within their respective far-IR luminosity class,
we find that FN1-40 lies midway into the first quartile of 60/100 values
for L$_{\rm FIR}\sim10^{11.6}$, while FN1-64 lies at the
first quartile point for L$_{\rm FIR}\sim10^{12.1}$. This is roughly
consistent with the extrapolated relation for the BGS shown in Figure~4. 
We discuss the implications of this discovery in section 4.

%
%
\begin{inlinefigure}
\centerline{\psfig{figure=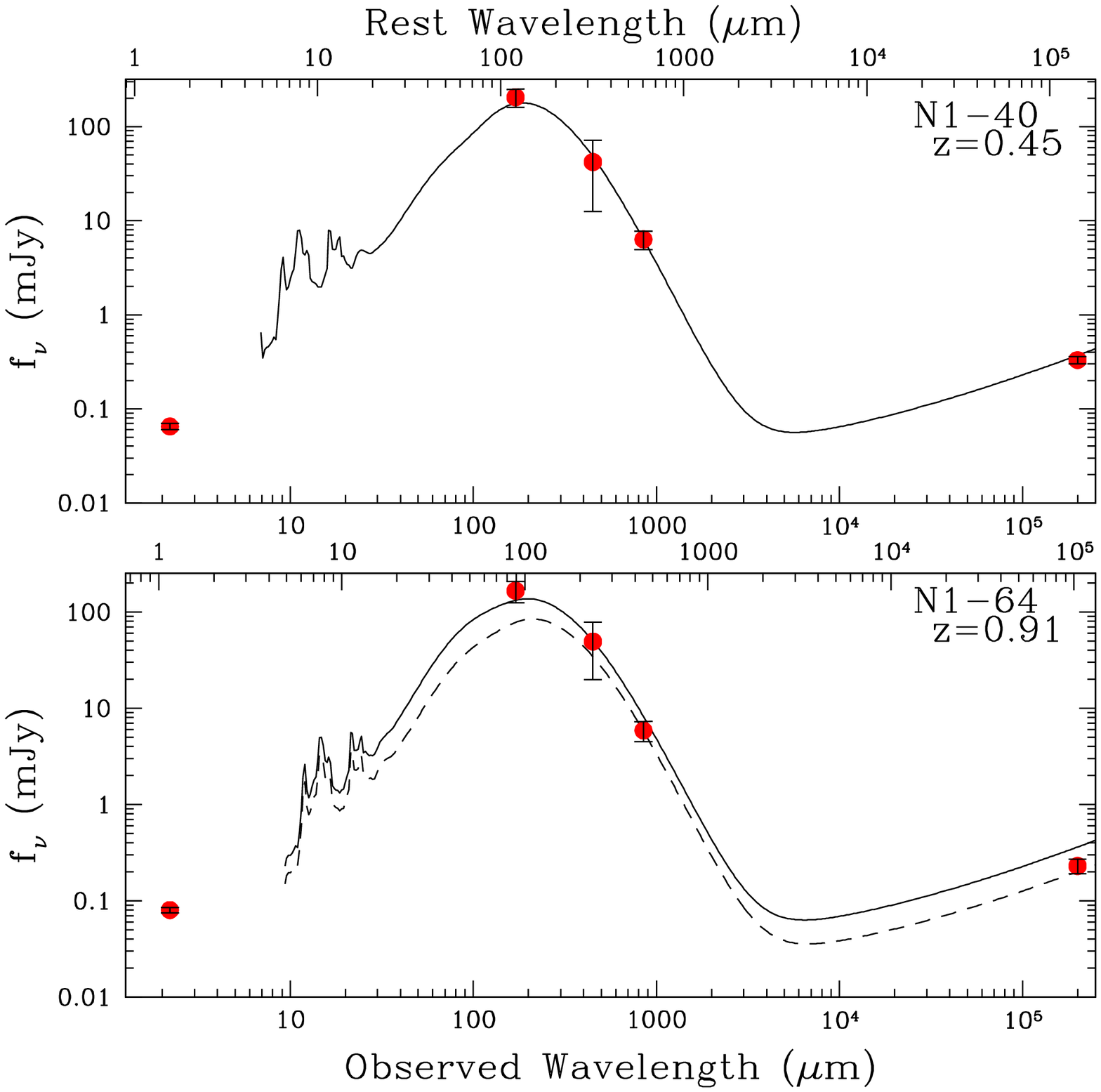,angle=0,width=3.5in}}
\figurenum{3}
\caption{\footnotesize
The observed spectral energy distributions of FN1-40 (top) and FN1-64
(bottom) from optical through radio measurements.  The best fit models
from the catalog of Dale et al.\ (2001) are overlaid, illustrating the
colder nature of FN1-40 at $z=0.45$. FN1-64 has hotter dust but is still
relatively cool compared to Arp\,220 (a ULIRG which has only 1/4 the bolometric
luminosity of FN1-64). For N1-64, a simultaneous fit to 170\,\m, 850\,\m, and
1.4\,GHz is rather poor, and an improved, cooler dust, 
fit to the 450\,\m, 850\,\m, 1.4\,GHz bands is shown as a dashed line
(motivated by the possibility of an additional source of far-IR emission
contributing to the large ISO-FIRBACK beam).
}
\label{fig3}
\end{inlinefigure}

%
%

\subsection{Optical properties and morphologies}


The existing optical imaging (1.2\arcsec\ FWHM) reveals  
an obviously merging, $r=21.4$ elongated source for FN1-64, 
and barely detected
$r\sim24.5$ components for FN1-40, Our $0.4''$ UKIRT imaging at $K$-band,
however, provides significant morphological information (Fig.~1).

The $K$ image of 
FN1-64 resolves into a pair of galaxies, with approximately equal
$K$-band flux, and the complete system has a $K=18.15$.  However, only
the northern $K$-band source has a bright optical counterpart, with
only a faint $r=23.5$ tail corresponding to the southern source.
Hence this pair  comprise a blue component to the north and a red
$(r-K)\sim5$ galaxy in the south.  The latter  is coincident with the
radio emission and by implication is probably generating the bulk of
the far-IR output. The galaxies are separated by $2.4''$ which
translates into 20\,kpc at $z=0.91$ and there is a bridge of emission
connecting the two components, suggestive of tidal debris.

FN1-40 is a faint galaxy ($r=24.5$),
with a possible close companion at
$2.3''$ separation (J2), and a third optically bright galaxy (J3) lying
$3.6''$ away from J1.  The radio centroid is coincident with component
J1, the $K$-brightest source with $(r-K)=4.8$, as labelled in Fig.~1.  
Our spectrum
identifies this as a galaxy at $z=0.45$. As the J3 source ($z=0.40$)
has a velocity offset of $\sim13000$\,km\,s$^{-1}$ from J1, it cannot
be considered a part of the system.  Both J1 and J2, however, show low
surface brightness extension (especially in the $z'$ and $i$ images)
along the  axis separating the two
components, and we suggest that this probably represents a merging
system.

A comparison of the morphologies of these two FIRBACK sources, multiple
components separated on scales of $\sim 10$\,kpc, with local ULIRGs
(Goldader et al.\ 2002) suggests a similarity with ``early stage
mergers'' such as VV\,114.  Equally, the tendency for  the radio
emission (and therefore presumably the far-IR) to come from the redder
component is also shared with local ULIRG samples.  However, there are
differences:  The far-IR curves for FN1-40/FN1-64 are fit by cool SEDs,
which would normally be accompanied by a modest L$_{\rm FIR}$/L$_{\rm optical}$
ratio for such early stage mergers (Dale et al.\ 2000). In contrast,
the measured L$_{\rm FIR}$/L$_{\rm optical}$ ratio, considering both the J1 and
J2 components in each case, is large (20.7 and 36.2) for FN1-64 and FN1-40 
respectively). 
L$_{\rm optical}$ is calculated from the integrated flux under a power law
fit to restframe wavelengths 0.5\mum\ to 1.5\mum. 
These values are similar to Arp\,220, a more evolved ULIRG with a hotter
dust temperature ($\sim50$\,K).

%
%
\begin{inlinefigure}
\centerline{\psfig{figure=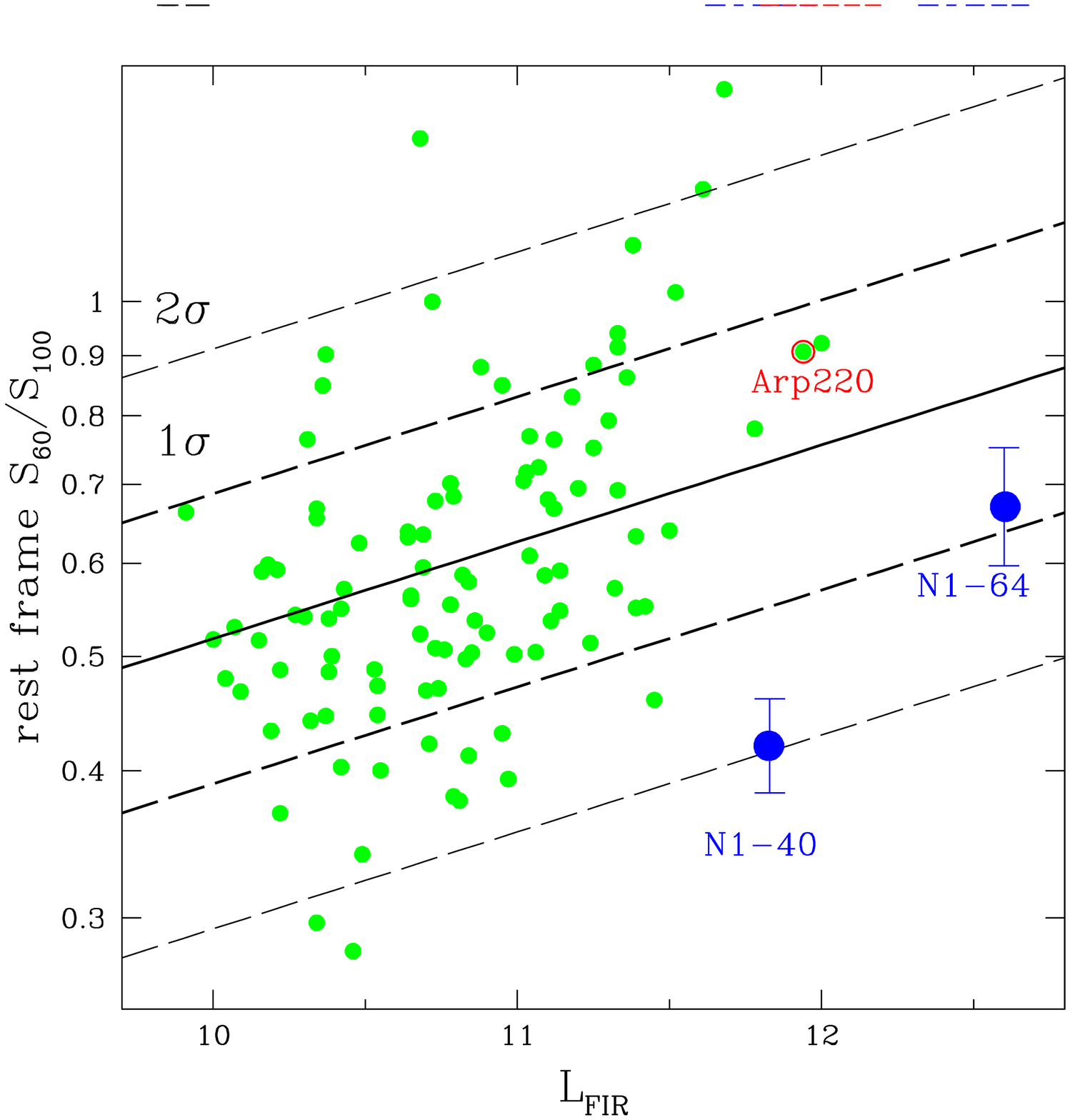,angle=0,width=2.9in}}
\figurenum{4}
\caption{\footnotesize
The dependence of rest frame 60/100 micron ratio (roughly dust temperature) 
with L$_{\rm FIR}$ for the Dunne et al.\ (2000) SCUBA observed sample
of BGS IRAS sources, 
compared to our two FIRBACK galaxies, with error bars denoting the 
model fitting error from photometric 
uncertainties. The correlation and 1$\sigma$, 2$\sigma$ 
deviations are overlaid. Arp220, the ULIRG typically used for comparison
with high$-z$ luminous galaxies, is identified.}
\label{fig4}
\end{inlinefigure}

%
%
\begin{table*}[hbt]
{\scriptsize
\begin{center}
\centerline{\sc Table 1}
\vspace{0.1cm}
\centerline{\sc Properties of FIRBACK FN1-64 and FN1-40}
\vspace{0.3cm}
\begin{tabular}{lll}
\hline\hline
\noalign{\smallskip}
{Property} &  {FN1-40} & {FN1-64} \cr
\hline
\noalign{\smallskip}
R.A.\ (J2000)	& 16:09:28.01	& 16:08:25.33\cr
Dec.\ (J2000)	& 54:28:32.6	& 54:38:09.5\cr
redshift$_{\rm spec}$     & 0.449$\pm$0.003   &    0.907$\pm$0.001 \cr
redshift$_{\rm phot}$     & 1.1$\pm$0.2   &    1.3$\pm$0.2 \cr
$U$    &  $>24.4$/$>24.4$  & 23.30/$>24.6$ (21.28)\cr
$g$    &  $>24.7$/$>24.7$  & 24.33/23.10 (21.91) \cr
$r$    &  24.45/24.57  & 23.48/22.52 (21.37) \cr
$i$    &  23.96/23.81  & 21.64/20.75 (19.93) \cr
$z'$    &  22.31/22.40  & 20.09/19.82 (18.83)\cr
$K$    &  19.73/20.51  & 18.63/18.71 (18.15) \cr
S$_{170\mu{\rm m}}$ (mJy)    &  $204.7\pm44.6$  & $166.2\pm42.0$ \cr
S$_{450\mu{\rm m}}$ (mJy)    &  $42.0\pm29.5$  &  $49.7\pm19.4$ \cr
S$_{850\mu{\rm m}}$ (mJy)    &  $6.3\pm1.4$  & $5.9\pm1.4$ \cr
S$_{20\,{\rm cm}}$ ($\mu$Jy)    &  $330\pm30$  & $230\pm40$ \cr
L$_{\rm FIR}^2$  & 6.7$\times$10$^{11}$  & 4.0$\times$10$^{12}$  \cr
L$_{\rm TIR}^3$  & 1.6$\times$10$^{12}$  & 7.7$\times$10$^{12}$  \cr
\noalign{\smallskip}
$[O{\sc ii}]$3729/3726  & 1.45 & 0.72 \cr
H$\alpha$/H$\beta$ & 9.54 & n/a \cr
$[O{\sc iii}]$5007 / $[N{\sc ii}]$6584 & 1.66  & n/a \cr
H$\gamma$/H$\beta$ & $<0.2$ & 0.24 \cr
$R_{23}$ & 6.27 & 3.56 \cr
$[O{\sc iii}]$4959+5007 / $[O{\sc ii}]$ & 0.61 & 0.95  \cr
SFR (H$\beta$) (reddening corrected) & 20 & 416 \cr 
SFR (far-IR) & 160 & 950 \cr
FIR/Opt$^4$ & 36.2& 20.7 \cr
60/100 (rest frame) & 0.43 & 0.69 \cr
\noalign{\hrule}
\noalign{\smallskip}
\end{tabular}

\begin{tabular}{cl}
$^1$ & Quantities with divisions are quoted for
components J1/J2 respectively. Magnitudes are \cr
{} & calculated
for 2\arcsec/2\arcsec\ apertures (for J1/J2). \cr {} &
Quantities in brackets are for the total system in the case of the obviously
merging FN1-64.\cr
{} & 3$\sigma$ photometry limits are ($U,g,r,i,z',K$)=(24.4,24.7,24.6,23.9,22.4,20.5). \cr
$^2$ &  S$_{\rm FIR}$ = (1.26$\times 10^{-14}$) * (2.58$\times$S$_{60}$ + S$_{100}$) $[$W/m$^2]$; 
	L$_{\rm FIR}$ computed using a flat, $\Lambda=0.7$ cosmology.\cr
$^3$ &  L$_{\rm TIR}$ covers 3\m\ to 1100\m\ as defined in Dale et al.\ (2001a).\cr
$^4$ & The {\it obscuration} ratio between FIR luminosity and optical/near-IR
luminosity, as described in the text.\cr 
\end{tabular}

\end{center}
}
\label{tab1}
\end{table*}

\subsection{Spectral Properties and Metallicities}

Combinations of strong emission lines can be used to determine or
constrain the nature of the ionizing radiation (Veilleux \& Osterbrock
1987), the amount of reddening to the emission region, and the
metallicity of the interstellar medium (ISM; e.g., Pagel et al.\ 1979;
Kobulnicky et al.\ 1999).  It is
important to stress that these measurements only reflect the
environment from which photons can escape and it is possible
that the most active regions within these systems are so highly
obscured that they do not contribute to the optical spectra used
in our analysis.  Moreover, in the case of FN1-64/J2, our optical
spectrum is offset from the very red object, radio emitting component of
the system (although this has not traditionally limited the analysis
of other similar systems, e.g.\ Ivison et al.\ 2001). 

The narrow lines and the resolution of the [O{\sc ii}] doublet in both
systems suggest no AGN component is present in either FN1-64/J2 or
FN1-40/J1, although without direct spectroscopic information from the
far-IR dominant region of FN1-64/J2, we cannot rule out an AGN origin
within this component (J1).  The optical spectra of both FN1-64/J2 and
FN1-40/J1 resemble that of luminous H{\sc ii} regions.

For FN1-40/J1, we can calculate the optical reddening based on the
Balmer ratio, H$\alpha$/H$\beta$. The region emitting Balmer lines is
apparently highly reddened, equivalent to an A$_V$=3.6 mag, and
consistent with the copious far-IR emission being centralized in the
same location as the optically brighter region for which the spectra
diagnostics have been measured. 
Applying this extinction to the H$\beta$ inferred star formation rate (SFR)
suggests a correction from 0.7\,M$_\odot$\,yr$^{-1}$ to 
20\,M$_\odot$\,yr$^{-1}$,
still well beneath that extrapolated from the far-IR 
(160\,M$_\odot$\,yr$^{-1}$).
We are able to probe the extinction in
FN1-64/J2 through the H$\beta$/H$\gamma$, revealing a more modest
$A_V=2.1$. 
As stressed before this constraint is not for the
major source of far-IR emission in this system (J1).
However, the H$\beta$ implied SFR from J2 (416\,M$_\odot$\,yr$^{-1}$), 
is comparable to the far-IR derived
SFR from J1 (950\,M$_\odot$\,yr$^{-1}$). 
The FN1-64 system is therefore reminiscent of other high$-z$ ULIRGs
where the optically bright component provides an insight into the
luminosity of the obscured portion (e.g.~Ivison et al.~2001 -- 
see discussion for further examples).

 
To measure the ISM metallicity, a rigorous determination requires
knowledge of electron temperature derivable only from intrinsically
weak lines. However, metallicity diagnostic line ratios based on
stronger lines have been empirically calibrated.  In particular, the
$R_{23}$ parameter, $R_{23}$ = ([O{\sc ii}]3727 + [O{\sc
iii}]4959+5007)/H$\beta$ (Pagel et al.\ 1979), has been 
calibrated against metallicity, with an intrinsic scatter of only 0.2
dex. It is a weak function of the ionization ratio [O{\sc
iii}]4959+5007/[O{\sc ii}]3727.  A reversal in $R_{23}$ occurs at $Z
\sim 0.3 Z_{\odot}$ due to cooling effects, so a low- and a
high-metallicity solution are associated with most values of $R_{23}$.
This degeneracy can however be broken using the [O{\sc iii}]5007/[N{\sc
ii}]6584 ratio (Kobulnicky et al.\ 1999).

The metallicity can be used as an indicator of
a galaxy's evolutionary state, and as a constraint on
possible present-day descendants.  Carollo \& Lilly (2001) have
investigated the ISM metallicities of H$\beta$-luminous field galaxies in the
$0.5<z<1.0$ redshift interval from the Canada-France Redshift Survey
(CFRS).  These galaxies fall in the $R_{23}$ versus [O{\sc iii}]/[O{\sc
ii}] plane in locations that are occupied by galaxies in the local
Jansen et al.\ (2000) sample. At both epochs, galaxies selected to have
the same H$\beta$ luminosities exhibit the same range of $R_{23}$ and
[O{\sc iii}]/[O{\sc ii}].
We find that the
galaxies FN1-40/J1 and FN1-64/J2 are within our 1$\sigma$
error of the median properties of the CFRS sources in the [O{\sc
iii}]4959+5007/[O{\sc ii}]3727 versus R$_{23}$ plane.  For the lower
redshift FN1-40 source, we can use the detected 
[N{\sc ii}6584] to break the metal
degeneracy of the R$_{23}$ indicator. 
The $[OIII]5007/[NII]6584 < 2$ necessitates 
metallicities $Z > 0.5 Z_\odot$, placing the source on the
upper branch of the relation, 
with a metallicity close to the solar value.

While the Carollo \& Lilly (2001) galaxies in the $0.5 < z < 1.0$ range
are bright star forming galaxies, they fall an order of magnitude
below the bolometric luminosity of our two FIRBACK sources under study
(several Carollo \& Lilly~2001 galaxies are detected in the 
CFRS 15\m\ ISO sample of Flores et al.~1999).
While it is therefore
interesting to compare R$_{23}$ metallicities for these objects, we
stress that the $R_{23}$ indicator has only been calibrated for low
obscuration regions, and may not be a valid surrogate for metallicity
in the highly obscured environments of ULIRGs.  In particular, we are
only measuring $R_{23}$ from regions which happen to be visible from a
system which is likely to be generally obscured.  


%
%
\section{DISCUSSION}

At $z<1$, sources observed at both 170 and 850\,$\mu$m will exhibit
similar characteristics as a function of dust temperature (Blain 1999);
the coldest and most dusty galaxies will have the greatest flux
densities for an equivalent far-IR luminosity.  Beyond $z\sim1$, the
170\,$\mu$m band lies blueward of the peak in the restframe black body
and dust temperatures T$_d<100$\,K no longer have significant effects
on the observed source flux.  A flux-limited survey such as FIRBACK is
thus expected to be biased towards cooler galaxies at higher
redshifts.  While we do not yet have the statistics to quantify the
actual selection rate, we note that the identification of the unusually
cold ULIRGs FN1-40 and FN1-64  are in agreement with this prediction:  we
find cooler galaxies than expected for the $\sim10^{12}$\,L$_\odot$
bolometric luminosities.

Models of the FIRBACK population (Dole et al.\ 2001) have
evolved the local luminosity function by boosting only the ULIRG
contribution with redshift, assuming an average SED template for
ULIRGs. This scenario is able to fit the FIRB and the FIRBACK counts,
inferring that 30\% of these sources lie at $0.5<z<1.0$, and 10\% at
$1<z<2.5$ (Dole et al.\ 2001).  However, the assumption that an average
ULIRG template can represent the FIRBACK source population is flawed as
it does not take into account the much colder sources which will be
preferentially selected at 170\m\ for $z<1$.
Other recent models of the far-IR galaxy population (Franceschini et al.~2001;
Chary \& Elbaz 2001) do not incorporate a cold luminous galaxy template
and cannot directly acount for numerous sources such as FN1-40/FN1-64.
Revisions to the Dole et al.\ model (Lagache et al.~2002)
include a population of colder luminous objects
based on the SED shape of the less luminous local FIRBACK sources.

The actual redshifts of
both FN1-40/FN1-64 ($z=0.45$/$z=0.91$) 
lie significantly below the $z=1.1$/$z=1.3$ predicted from
the submm and radio, using the Carilli \& Yun
(2000) estimator (with 1$\sigma$ uncertainty of 0.17) 
which is also derived from the average ULIRG SED. 
Moreover, if the majority of FIRBACK sources were to span the range in
properties observed in these two galaxies, the average properties deduced
from an Arp220 template would be severely skewed to higher redshifts.
We chose these two sources for Keck follow-up at random from the 
FIRBACK sources where the sub-mm/radio ratio suggested a high ($z>1$) redshift.
As a consequence, the distinct possibility exists that almost none of
the radio identified 
FIRBACK sources may lie at $z>1$, dependent on the mid- and
high-redshift distributions in dust temperature.

The measurement of the 850/1.4 ratio for both these sources does allow
a constraint on its use as a redshift indicator.  Clearly the ratio for
these sources must lie significantly above the canonical value for
ULIRGs found by Carilli \& Yun (2000), as shown in Fig.~5.  The
existence of such cooler luminous sources may affect seriously the
interpretation of the high redshift SCUBA selected population
(see also Eales et al.~2000).

Cold yet luminous dusty sources suggest large masses of dust heated at
moderate intensity, rather than small amounts of dust in a compact
configuration subjected to extremely intense radiation fields.  This latter
scenario is clearly supported by the spatial and SED properties of the nearby
ULIRG Arp220 (Soifer et al.~1984).  
While one could invoke evolving or novel dust properties to explain cooler
high-luminosity sources like FN1-40 and FN1-64, there is no evidence to
support such unusual dust properties, especially that the SEDs of both
FN1-40 and FN1-64 are fitted quite well with the Dale et al.\ models from the
local Universe.
For objects like FN1-40 and
FN1-64 however, the data suggest a spatially distributed star formation
activity heating a more extended ISM.
Arp 302 represents a nearby system with a 60/100 color as cold as FN1-40
and L$_{\rm FIR}$=5$\times$10$^{11}$\,L$_\odot$. Lo, Gao \& Gruendl (1997)
have shown that luminous CO emission traces the entire optically defined
disks of both merging components, covering 23\,kpc and 11\,kpc respectively,
and we might imagine a similar configuration in FN1-40. 
The main difficulty then is to explain the high extinctions suggested by the
large infrared-to-visible ratios in these systems.  Since we know little of
their detailed morphology or geometry, we cannot exclude the possibility
that the initial conditions and the details of the encounters (strong tidal
interactions rather than direct collisions or mergers) result in obscured
yet distributed rather than strongly concentrated star formation activity.

%
%
\begin{inlinefigure}
\centerline{\psfig{figure=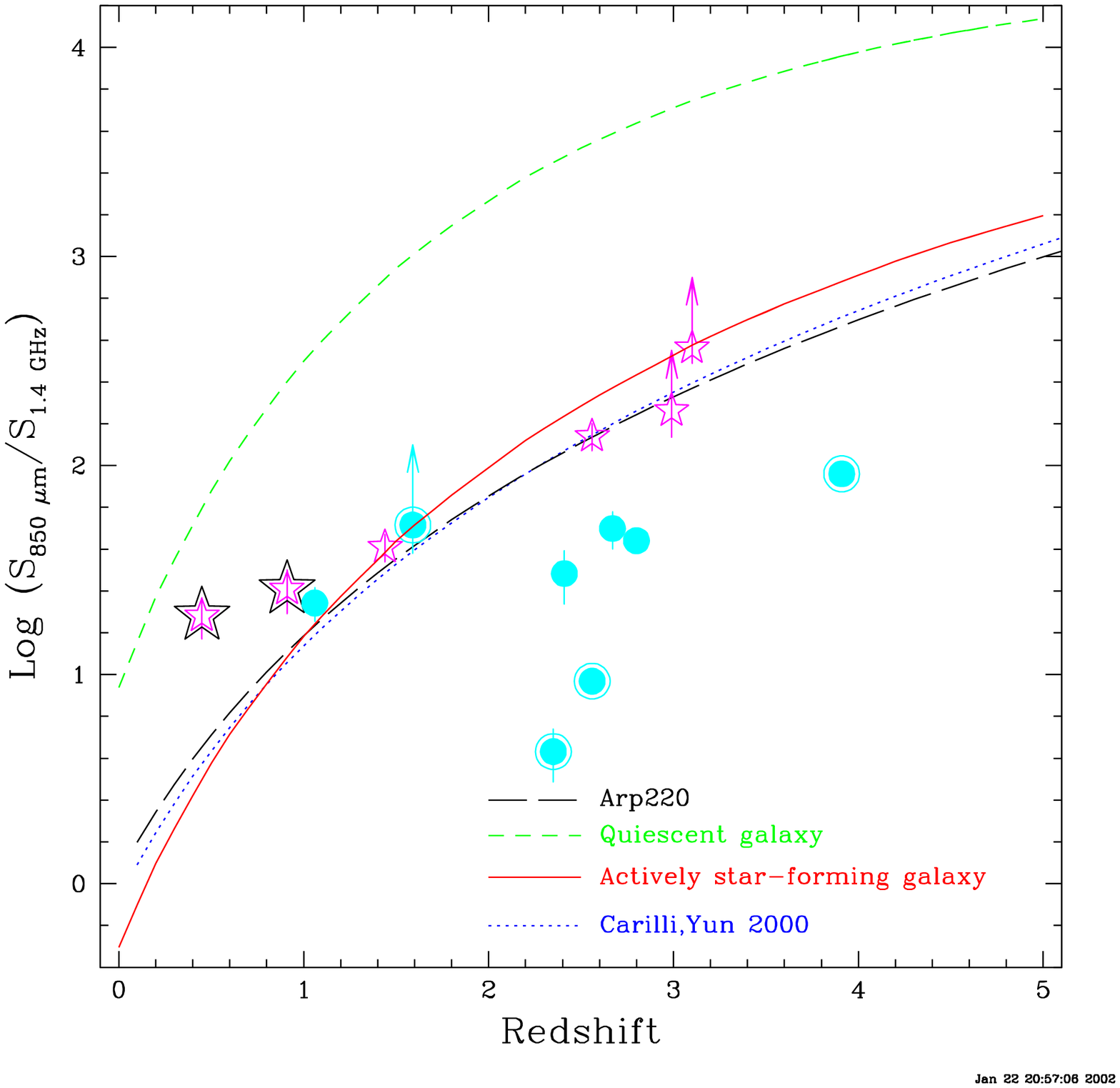,angle=0,width=2.8in}}
\figurenum{5}
\caption{\footnotesize
The 850\m/1.4\,GHz ratio as a function of redshift for galaxies of various
activity levels, ranging from normal spirals to starbursts, with the
Carilli \& Yun (2000) relation shown for ULIRGs.
Our two cold FIRBACK sources are shown double stars. Other star symbols 
depict sub-mm sources with no sign of AGN activity (in increasing 
redshift order: HR10--Dey et al.\ 1999, SMM\,J14011+0252--Ivison et al.\ 2001,
MMD11--Chapman et al.~2002b, SA22-blob1--Chapman et al.~2001).
For comparison we plot 4 sub-mm selected sources identified as AGN
with filled circles (Smail et al.\ 2002) and 4 BAL quasars which were
sub-mm detected (double circles -- Lewis \& Chapman 2002).}
\label{fig5}
\end{inlinefigure}

\subsection{Comparison with higher redshift SCUBA populations
} 

Figure~\ref{fig5} shows how these FIRBACK sources compare to other submm
luminous sources with measured spectroscopic redshifts.  Note that while the
FIRBACK sources appear to be pure starburst, there are very few
spectroscopically identified sub-mm galaxies at high redshift for
comparison which do not show AGN signatures
(SMM\,J14011+0252 at $z=2.56$ -- Ivison et al.\ 2001,
Westphal~MMD\,11 at $z=2.99$ -- Chapman et al.\ 2002b,
HR\,10 at $z=1.44$ -- Dey et al.\ 1999).
Both FN1-64 and FN1-40 appear morphologically similar to the SCUBA
galaxies SMM\,J14011+0252 and Westphal~MMD\,11, having a red $(R-K)$ source
identified to the far-IR emission along with a bluer companion.  This is
similar to several local ULIRGs (e.g.\ 08279+1372 -- Surace \& Sanders
2000; Goldader et al.\ 2002), and may suggest that all of these galaxies
are generating their far-IR output by similar mechanisms.

As these two sources lie midway between the local {\it IRAS}
population, and the high redshift SCUBA-selected population
(e.g.\ Smail et al.\ 2002), understanding their properties provides a
crucial benchmark for continued progress on the evolution of
IR-luminous galaxies.  FN1-40 and
FN1-64 are representative of the optically faint FIRBACK sample with 
radio and SCUBA
detections ($\sim$30\% -- Scott et al.\ 2000), 
comprising $\sim$8\% of the radio identified
FIRBACK survey.  In terms of the
bright (S$_{850 \rm \mu m}>6$\,mJy), blank field submm sources (e.g.\ Smail et al.\ 2001, Scott et al.\ 2001) in this area on
the sky, they would represent only a small fraction of the 850\,\m\ counts
($\sim3$\%).

The strong bias to colder dust temperatures in sub-mm and far-IR selected 
samples occurs because the selection bands fall redward of the grey-body peak.
At an observed 170\,\m, this implies the cold bias out to $z\sim1$
discussed earlier.
However, for 850\,\m, this bias is inherent in the entire plausible range of 
galaxy evolution, out to $z\sim8$ (e.g.~Blain\ 1999).
In this sense, the wide field SCUBA surveys will preferentially
uncover the coldest sources for a given luminosity.

We would like to use the existence of these cold, luminous sources to 
place a constraint on the dust temperature distribution at high redshifts.
If the important driver of the dust temperature is the radiation
field then we must compare at similar luminosity systems.
Figure~4 suggests that these luminous FIRBACK sources are within
the coldest 16\% of the local population if we cut at the same luminosity.
If we propose that this local population of IR-luminous objects undergoes
a strong luminosity evolution of the form $(1+z)^4$ out to $z\,{=}\,1.5$,
we can determine the relative numbers of cold and warm objects 
isolated by our selection function.
This form of evolution has been shown to match
the observe space density of high$-z$ ULIRGs to the local population
(e.g.~Blain et al.~1999; Chapman et al.~2002c).
 
We start with the FIRBACK survey limit, S$_{170} > 120$\,mJy.
We add the requirement that S$_{850}$/S$_{1.4}$ give 
${(1+z)}\over{T_{\rm d}/50\,K}$ $ > 2$, the Carilli \& Yun (2000) criterion
we employed to select high redshift objects for spectroscopic followup
with Keck. This selection function implies that we begin to detect objects
with T$_{\rm d}<30$\,K at $z>0.35$, and T$_{\rm d}<50$\,K at $z>1$.
We find 35\% more FIRBACK detectable objects with 
T$_{\rm d}<30$\,K than with 30\,K$<T_{\rm d}<50$\,K in the 
accessible volume from $z=0.35$ to $z=1.6$, where we can no longer
detect a $50$\,K source.
As our two detected sources have T$_{\rm d}$=26\,K and 32\,K, our 35\%
bias towards selecting T$_{\rm d}<30$\,K sources implies that the outcome
of our experiment deviates by less than $1\sigma$ from the expected outcome.

We can conclude that the dust temperature distribution at higher redshifts is
consistent with
that observed locally (or alternatively, that the characteristic temperature
of a ULIRG doesn't increase dramatically out to $z\sim1$).
Note that this is not necessarily the expected result.
For instance, different temperatures distributions for similar
luminosity ULIRG populations at $z=0$ and $z=1$ could arise due to 
different geometry of the active region, or different chemical properties of 
the dust.
Complete redshift distributions for the FIRBACK population will be 
required to carefully test the dust temperature distributions at low and
high redshift.
A more detailed model,
taking acount of the local bivariate luminosity function in
L$_{\rm TIR}$ and S$_{60 \rm \mu m}$/S$_{100 \rm \mu m}$, studies directly
the effect of the complete distribution of dust properties on the high
redshift far-IR population (Chapman et al.~2002a).

\section{CONCLUSIONS}
We have obtained Keck spectra and UKIRT high spatial resolution near-IR
imagery for two of the proposed highest redshift sources from the
FIRBACK-N1 170\m\ survey. This survey is currently the most
sensitive probe of the properties of dusty galaxies at the peak of the
FIRB.

We find that the redshifts of counterparts to the 170\,\m\ sources
confirm that both sources are ULIRGs, but that their redshifts are
significantly lower  than implied by fitting a typical ULIRG SED to
their far-IR/submm/radio SEDs.  This indicates that they have cooler
dust temperatures, T$_d\sim30$\,K, than the canonical ULIRG values
(T$_d\sim50$\,K).  The sources both show morphologies suggestive of
early stage mergers, similar to recently identified SCUBA galaxies, and
also many local ULIRGs.

For FN1-40, we are able to examine optical line diagnostics in the
region of far-IR emission. We measure a large, A$_{\rm V}\sim4$\,mag,
extinction, indicating a large correction to the H$\beta$ estimated
SFR. The metallicity in this region, as probed by the R$_{23}$ and
indicator and OIII/NII ratio, is found to be typical for starforming
galaxies  2 orders of magnitude less luminous (close to solar).

The detection of these galaxies in the submm by SCUBA allows a
constraint on the 850/1.4 redshift indicator. We find a relation for
these galaxies lying significantly above the relation found by Carilli
\& Yun (2000), due to their cooler dust temperatures.  The existence of
such cooler, but still luminous sources may thus affect the
interpretation of the SCUBA selected population,
if they are numerous at high redshift.  
Recent models of the far-IR galaxy counts, relying on the rapid
evolution of a hot ULIRG component to the local luminosity function,
may predict an erroneous redshift distribution without taking into
account sources such as FN1-40 and FN1-64.

We have turned the discovery of these two cold, luminous galaxies
into an estimate for the dust temperature
distribution at $z\sim1$ for the ULIRG population. We conclude 
that our 170\mum, 850\mum, and 1.4\,GHz selection function isolates
approximately 35\% more cold sources ($<30$\,K) than warmer sources
($30<$T$_{\rm d}<50$) over the detectable redshift range,
under the assumption that the temperature distribution at $z=0$ and $z=1$ is
similar.
The relative proportion of the cold
population at low/high-$z$ can therefore be understood as consistent
with the local 60\mum/100\mum\ distribution.

\acknowledgements
Thanks to A.\ Blain for stimulating conversations about this source population.
We acknowledge the useful comments of an anonymous referee, which helped improve
the text.
The INT WFC data was made publically available through the Isaac Newton Groups'
Wide Field Camera Survey Programme. The Isaac Newton Telescope is operated
on the island of La Palma by the Isaac Newton Group in the Spanish
Observatorio del Roque de los Muchachos of the Instituto de Astrofisica de
Canarias.
UKIRT is operated by the Joint Astronomy Centre
on behalf of the U.K. Particle Physics and Astronomy Research Council.
IRS acknowledges support from the Royal Society and the Leverhulme Trust.

\end{document}